# Segregation–Assisted Spinodal and Transient Spinodal Phase Separation at Grain Boundaries

Reza Darvishi Kamachali*,[1], Alisson Kwiatkowski da Silva[1], Eunan McEniry[1], Dirk Ponge[1], Baptiste Gault[1,2], Jörg Neugebauer[1], Dierk Raabe[1]

[1] *Max-Planck-Institut für Eisenforschung GmbH, Max-Planck-Straße 1, 40237 Düsseldorf, Germany*

[2] *Department of Materials, Royal School of Mines, Imperial College, Prince Consort Road, London SW7 2BP, United Kingdom*

Segregation to grain boundaries affects their cohesion, corrosion and embrittlement and plays a critical role in heterogeneous nucleation. In order to quantitatively study segregation and phase separation at grain boundaries, we derive a density-based phase-field model. In this model, we describe the grain boundary free energy based on available bulk thermodynamic data while an atomic grain boundary density is obtained using atomistic simulations. To benchmark the performance of the model, we study Mn grain boundary segregation in the Fe–Mn system. 3D simulation results are compared against atom probe tomography measurements. We show that a continuous increase in the alloy composition results in a discontinuous jump in the Mn grain boundary segregation. This jump corresponds to an interfacial spinodal phase separation. For alloy compositions above the interfacial spinodal, we found a transient spinodal phase separation phenomenon which opens opportunities for knowledge-based microstructure design through the chemical manipulation of grain boundaries. The proposed density-based model provides a powerful tool to study thermodynamics and kinetics of segregation and phase separation at grain boundaries.

*Corresponding Author, Email: kamachali@mpie.de

# 1   Introduction

Grain boundaries (GBs) influence functional and structural properties of polycrystalline materials. They can have both positive (strengthening in terms of the Hall-Petch effect) and negative (preferred site for corrosion and decohesion) influence on the material's performance. Hence, studying and engineering GBs are crucial for optimizing microstructure and material properties [1, 2]. The vast structural variability and high amenability of GBs to chemical changes renders them ideal objects for tuning their properties by solute segregation [3–7]. From a thermodynamic point of view, a GB has distinct phase-like behavior [8] which is a function of the relevant thermodynamic state variables. Consequently, GBs may undergo confined structural and chemical transitions as evidenced in several systems [9–11]. Compared to bulk materials, a GB is subjected to additional (geometric) constraints that result in structural and compositional gradients within the GB region. GB phases are hence sometimes referred to as complexions [11–13] to distinguish them from bulk phases which are assumed to be homogeneous by definition.

In alloys, the properties of GBs are expected to closely correlate with their composition. This becomes particularly important when solute atoms segregate to the GBs. The segregation is thermodynamically driven by a reduction in the total energy (GB energy plus bulk energy) of the system. Segregation not only alters the local kinetics [14–19] and mechanical properties of a GB [20–23], but also influences thermodynamic driving forces for heterogeneous nucleation such as in GB premelting [24, 25], phase transformations [26–28] and precipitation in different alloys [29–33]. Before reaching saturation, segregation can also result in GB phase separation forming solute-poor and solute-rich regions inside the GB plane. This has been indeed postulated by Fowler and Guggenheim [34] and Hart [35, 36] who suggested that an interface/GB region exhibiting similar properties to a regular condensed solution might undergo such a phase separation. The resulting low-dimensional (planar) spinodal phase separation into high and low segregation regions produces precursor states for the formation of new phases [37, 38]. These precursor states are either confined to the GB or expand as a regular volume phase into the adjacent bulk.

In reality, one expects segregation, phase separation and nucleation of a new phase at a GB to occur hand in hand. In order to describe simultaneous segregation, phase separation and their impacts on microstructure evolution, the thermodynamics and kinetics of GB segregation must be studied quantitatively. Extensive efforts have been made to apply surface adsorption models [8, 34, 39, 40] for understanding GB segregation [35, 41–45]. Assessing the GB thermodynamics and kinetics is, however, a challenging task due to the complex nature of GBs [1]. To address this problem, several models have been developed. For example, to account for changes in thermodynamic properties of GB, variations in the local coordination number (broken bond model) of the defected GB structure were considered (for details see [46–48] and references therein). In a recent study, we have employed this concept to investigate GB segregation in magnetic Fe–Mn alloys [38].

In parallel, continuum phase-field models have been developed to study GB-related phenomena which concentrate on the effect of gradient energy terms within the GB region. The seminal work of Cahn on wetting [49] highlighted the importance of these gradient energy terms in the vicinity of an interface and their effect on the critical wetting transition. Since then several phase-field studies were conducted on the thermodynamic and kinetic effects of solute segregation [50–54]. Hu and Chen [55] developed a phase-field model for studying solute segregation and phase transition at dislocations. Ma *et al.* combined Cahn's non-local concentration gradient model with the idea of variation in local coordination number to describe GB transition and drag forces [56] as well as segregation to dislocations [57]. A further extension of their work has been presented recently [58]. Based on the Kobayashi, Warren and Carter (KWC) model for GBs [59], Tang *et al.* developed a phase-field model that describes order-disorder transitions in pure systems [60, 61] and phase transition in binary alloys [62]. Kim *et al.* [63] reviewed thermodynamic aspects of phase-field models for GB segregation. They proposed a phase-field model for GB segregation which uses the equal diffusion potential condition [64] (instead of the common equal composition condition). Compared to the previous phase-field models, Kim *et al.* [63] have shown that the new model can simulate a broader range of segregation in alloys especially when strong reductions in GB energy occur.

In the current study, we introduce a continuous atomic density field to approximate GB free energy functional based on available bulk thermodynamic data. In this picture, the structurally defected GB is represented by its lower average density with respect to the corresponding bulk. In order to determine the GB density profile, atomistic simulations were conducted. We describe our model in Sec. 2. The Supplementary Information provides detailed derivations of our mesoscale density-based free energy description. A coarse-graining scheme is discussed to link the atomistic simulations and the mesoscale model. We study here the Fe–Mn system in which the segregation of Mn in BCC Fe plays a critical role in determining GB mechanical properties and subsequent formation of austenite [27, 28]. ThermoCalc databases were used to obtain thermodynamic and kinetic information for the Fe–Mn system. In Sec. 3, assessment of the model parameters, the simulation results and the experimental studies are presented. Our study reveals that a transient spinodal phase separation can occur for a range of alloy compositions, while the equilibrium segregation isotherms indicate only a single critical bulk composition related to the GB spinodal. As a result, a consistent explanation of GB segregation behavior in three Fe–Mn alloys, namely Fe–3.0at.% Mn, Fe–4.0at.% Mn and Fe–8.6at.% Mn, is discussed. The existence and significance of a segregation-assisted transient spinodal phenomenon and its potential application for designing desirable microstructures are discussed in Sec. 4.

## 2 A Density-based Model for Grain Boundaries

In a GB, atoms are forced to accommodate for the incompatible lattices of the two adjacent grains. This results in a different atomic density in the GB when compared to the corresponding bulk values. In his seminal work, van der Waals [65] showed that the energy of an interface can be described as a function of mass density and its variations (gradients) within the interface region. In contrast to Gibbs's model [8] which assumes a mathematically sharp interface, in this picture an interface is a diffuse domain across which the density profile varies continuously albeit very sharply. For a GB, one expects such a gradient term to be confined to the GB region. Based on the van der Waals model, the Gibbs free energy density of a GB in a pure substance made of atomic species $A$ can be described as

$$G^{GB}(T,\rho) = E_A^B \left(\rho^2 - 1\right) + C_A^B(\rho - 1) + \frac{\kappa_\rho}{2}(\nabla\rho)^2. \tag{1}$$

The derivation of Eq. (1) is given in the Supplementary Information of this study. The superscripts $B$ and $GB$ represent the bulk and GB properties, respectively. Here $\kappa_\rho$ is the density gradient coefficient, $E_A^B$ is the potential energy of the bulk phase, which is a function of the atomic arrangement in the system, and $C_A^B = K_A^B + pV_A^B - TS_A^B$ is the sum of kinetic ($K_A^B$), mechanical work ($pV_A^B$) and entropic ($-TS_A^B$) energy contributions of the same material at a given temperature $T$. Neglecting the $pV_A^B$ term and the kinetic energy for a solid material, we can write

$$C_A^B \approx -TS_A^B. \tag{2}$$

In 1, we use a relative (dimensionless) atomic density filed $\rho$, continuously varying across the system, such that $\rho = \rho^B = 1$ inside the bulk (far from the GB) and $\rho < 1$ inside the GB region. In the center of the GB, $\rho = \rho^{GB} < 1$ marks the minimum GB atomic density at the GB plane. In principle the atomic density within a GB plane fluctuates. This is, however, neglected in the current treatment, assuming an average constant GB density value $\rho^{GB}$ corresponding to the GB type. The GB density value is calculable using atomistic simulations. As it is shown in the Supplementary Information, the relative atomic density field $\rho$ can be related to the atomistic description of the system by choosing a coarse-graining length comparable to the cut-off radius in the atomistic simulations.

The potential energy and entropy terms in Eq. (1) turn out to scale differently with the atomic density parameter $\rho$. The $-TS_A^B$ term scales linearly with the atomic density. This represents the change in the amount of materials (number of atoms per unit volume) which differs between the bulk and GB. The potential energy, however, scales quadratically with the density. In this case, the extra density coefficient is proportional to the bonding energies (force density) in the elastic limit. The gradient term in Eq. (1) is a correction to the potential energy due to the spatial density change within the GB region. Equation (1) allows an approximation of the GB free energy based on the bulk thermodynamic data and a relative density parameter. For more information see the Supplementary Information. In order to extend the model to a binary system, mixing enthalpy and entropy terms have to be introduced. The main contribution to the mixing entropy $\Delta S_{mix}$ is the configurational entropy due to the mixing of the solvent ($A$) and solute ($B$) atoms. For the sake of simplicity, we neglect the scaling of the mixing entropy with the atomic density. The enthalpy of mixing $\Delta H_{ex}$, however, may be strongly influenced by the chemical and structural environment. Motivated by the first term in Eq. (1), one can write

$$\Delta H_{ex} = \rho^2 \Delta H_{ex}^B \qquad (3)$$

in which $\Delta H_{ex}^B$ is the excess enthalpy of the bulk solution. For the Fe–Mn system, this will be the sum of chemical and magnetic mixing enthalpies. Using Eqs. (1)–(3), the total Gibbs free energy density of a heterogeneous binary system (containing a GB) can be written as

$$\begin{aligned} G_{\text{alloy}}(T, \rho, X_B) &= G^B(T, X_B, \rho = 1) + G^{GB}(T, \rho) \\ &= X_A G_A^0 + X_B G_B^0 + \rho^2 \Delta H_{ex}^B - T \Delta S_{mix}^B \\ &+ \frac{\kappa_X}{2}(\nabla X_B)^2 + E_A^B\left(\rho^2 - 1\right) - T S_A^B(\rho - 1) + \frac{\kappa_\rho}{2}(\nabla \rho)^2 \end{aligned} \qquad (4)$$

where the subscripts $B$ and $A$ indicate solute (Mn) and solvent (Fe), respectively, $X_B$ is the solute concentration field with $X_A + X_B = 1$, $G_i^0$ is the Gibbs free energy of the pure bulk $i$ and $\kappa_X$ is the concentration gradient coefficient. It is worth noting that a difference between solute and solvent atoms can result in the composition dependence of atomic density. For simplicity, however, this is neglected in the current formulation. The gradient coefficients $\kappa_\rho$ and $\kappa_X$ in Eq. (4) are obtainable from atomistic simulations and bulk thermodynamic databases, as discussed in the next section. For any point in the GB with atomic density $\rho < 1$, Eq. (4) gives an approximation of the GB Gibbs free energy density. Inside the homogeneous bulk phase, $\rho = 1$, $\nabla \rho = 0$, $\nabla X_B = 0$ and the Gibbs free energy of the bulk can be recovered from Eq. (4):

$$G_{\text{alloy}}^B = X_A G_A^0 + X_B G_B^0 + \Delta H_{ex}^B - T \Delta S_{mix}^B. \qquad (5)$$

The current model is a generalization of the previous work by Ma *et al.* [56] where the atomic density field can be considered as a coarse-grained coordination number on the continuous level. Although the current density-based model considers a single concentration field across the system, i.e. an equal composition condition as discussed by Kim *et al.* [63], here the density and concentration fields have two septate gradient energy contributions. While the density field governs the GB width and its dynamics, the gradient of the concentration field regulates the tendency and magnitude of the GB segregation and its possible spinodal decomposition [66]. The current model offers an approach to approximate GB thermodynamic functions based on available bulk thermodynamic data. In order to study isothermal GB segregation and phase separation, the time evolution of the concentration and density fields are calculated according to

$$\dot{X}_B = -\nabla \cdot \mathbf{J}_B = \nabla \cdot \left[ M X_B \, \nabla \frac{\delta \mathcal{G}}{\delta X_B} \right] \quad (6)$$

and

$$\dot{\rho} = -L \frac{\delta \mathcal{G}}{\delta \rho} \quad (7)$$

respectively, in which $\mathcal{G} = \int G_{\text{alloy}} dV$, $M$ is the concentration-dependent atomic mobility, $\delta$ indicates functional derivatives and $L$ is a positive mobility factor. As a diffusion-controlled process, the kinetics of segregation is governed by the mobility of Mn solute atoms in Eq. (6). The density field evolves much faster by atomic bond relaxation. To assure a diffusion-controlled kinetic, a large value for $L$ compared to $M$ is chosen (Table 1). The atomic mobility of Mn in the Fe–Mn system is extracted from the ThermoCalc MOB04 database. The thermodynamic functions required for Eq. (4) are extracted from ThermoCalc TCFE9 databases. For more information about the thermodynamic and kinetic databases see [67–72]. In order to obtain realistic values for the GB density and energy $\rho^{GB}$, atomistic simulations have been performed for a $\Sigma 9\{122\}[1\bar{1}0]$ symmetric tilt GB in $\alpha$-Fe, using the environmental tight-binding approach described in [73, 74]. Further details are described in the Methods section.

# 3 Results

## 3.1 Assessment of the model parameters

Since the current methodology can be applied for studying GB segregation and related phenomena in different materials, the associated model parameters are discussed to provide guidance for future studies.

**Grain boundary energy and atomic density profile**. For a flat GB in a pure substance, the equilibrium atomic density profile across the GB follows:

$$\rho_{eq}(x) = \begin{cases} \left(\frac{1+\rho^{GB}}{2}\right) - \left(\frac{1-\rho^{GB}}{2}\right)\cos\left(\frac{\pi x}{\eta}\right) & \text{if } -\eta \leq x \leq \eta \\ 1 & \text{else} \end{cases} \quad (8)$$

$$\text{with} \quad \eta = \pi\sqrt{\frac{\kappa_\rho}{-2E_A^B}} \quad (9)$$

which satisfies $2E_A^B\rho + K_A^B - TS_A^B - \kappa_\rho\nabla^2\rho = 0$ with $\rho(x=0) = \rho^{GB}$, $\rho(x=\pm\eta) = 1$ and the continuity condition $\nabla\rho(x=0) = 0$. Previous phase-field models for GBs, such as the KWC model [59], result in a discontinuous order-parameter at the GB. The current model allows for a continuous atomic density profile across the GB (see Figure 1). This is achieved since the current derivation of GB free energy Eq. (1) has three independent terms while conventional phase-field models for GBs contain two independent coefficients.

Here $2\eta$ is the GB width in the pure substance. Experimentally, $2\eta$ can be larger than the measurable values for GB width because the atomic density is a smooth and continuous field across the GB. Choosing the coarse-graining radius equal to the cut-off radius, as applied in the theories of atomistic simulations [75], we can link the density-based model parameters to the atomistic simulations. Inserting Eq. (8) in Eq. (1) gives the equilibrium GB energy $\gamma_A = \alpha_0\left(1-\rho^{GB}\right)^2$ with $\alpha_0 = \frac{\pi}{4}\sqrt{-2E_A^B\kappa_\rho}$. One can see that $\gamma_A$ varies as a function of $\rho^{GB}$. For a specific GB atomic density $\rho^{GB}$, the GB energy is obtained by determination of the parameters $E_A^B$ and $\kappa_\rho$. For $\rho^{GB} \to 1$ the GB energy approaches zero. This situation can indeed be observed for the case of special, highly symmetric boundaries, such as for coherent twin boundaries with low coincidence values, where the GB energy is low and the local atomic density is close to that of the bulk.

**Atomistic simulations of a grain boundary in $\alpha$-Fe.** The bulk potential energy $E_A^B$, the minimum atomic density $\rho^{GB}$ at the GB, gradient coefficient $\kappa_\rho$ and the GB energy can be determined directly from atomistic simulations. In the current study, a symmetric tilt GB $\Sigma 9\{122\}[1\bar{1}0]$ in $\alpha$-Fe is simulated. The current assessment, however, can be generalized and applied for different types of GBs. The GB structure is fully relaxed using an environmental tight-binding approach for Fe, a methodology previously used to study light-element interactions with a broad set of GBs in $\alpha$-Fe [73].

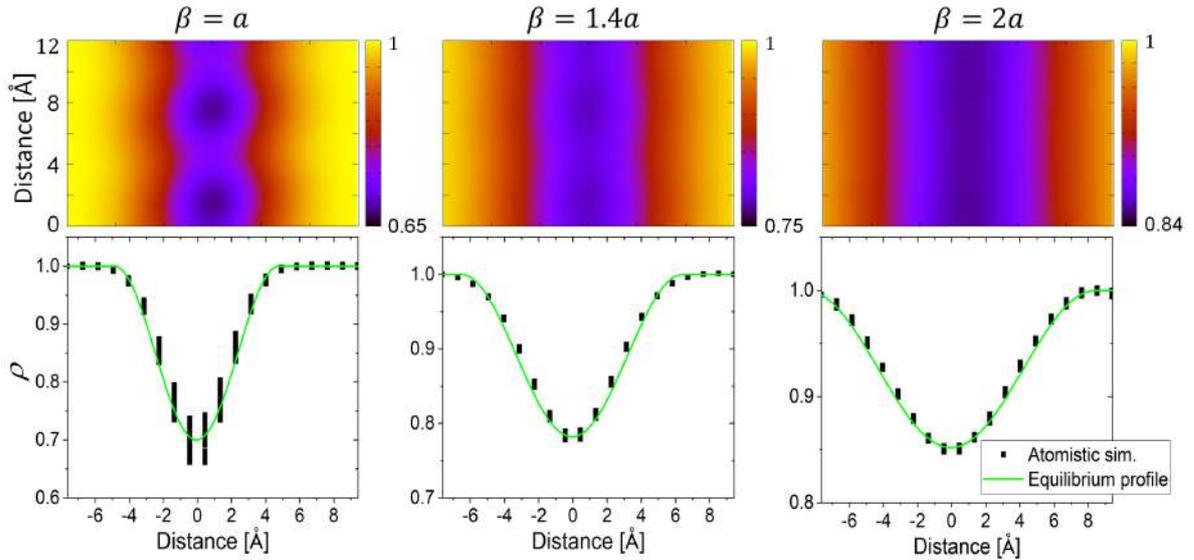

Figure 1: Density profile across a GB in $\alpha$-Fe. The results of atomistic simulations are shown and fitted to the analytical solution shown in Eq. (8) (solid line). A physically sound coarse-graining radius is equal to the cut-off radius of $1.4a$. Nevertheless, three different smearing radii $\beta = a, 1.4a$ and $2a$ (with the interatomic distance $a \sim 2.5$ Å) were examined to compare the results. See Methods section and Supplementary Information for more details.

In order to obtain the continuous atomic density profile from atomistic simulations a Gaussian broadening scheme with various smearing radii $\beta$ was applied. As discussed in the Supplementary Information, a physically sound choice of the coarse-graining length is to calculate the continuous atomic density over a spherical volume of the atomistic cut-off radius. This is motivated by the fact that all physical properties from atomistic simulations are calculated up to the cut-off radius. Three different smearing radii close to the cutt-off radius ($\sim 4$ Å) were examined in our study. These are illustrated in Fig. 1. With increasing the smearing radius, the atomic density profile becomes smoother and the in-plane fluctuations of the GB density decrease. By fitting the analytical solution given in Eq. (8) to the results from atomistic simulation, the corresponding values of the minimum GB atomic density $\rho^{GB}$ and the GB width $\eta$ can be obtained. The atomistic simulations confirm the continuity of the coarse-grained atomic density field $\rho$ across the GB which is obtained based on the mesoscale density-based free energy density Eq. (1).

**Gradient concentration coefficient** $\kappa_X$. In order to obtain the composition gradient coefficient $\kappa_X$, the interface between the spinodally decomposed low- and high-concentration bulk phases must be studied. As proposed by Cahn and Hilliard

[66], the energy of this interface is

$$\gamma_I = \sqrt{2\kappa_X} \int_{X_{low}}^{X_{high}} \sqrt{\Delta G_{\text{alloy}}^B}\, dX. \tag{10}$$

Here the subscript $I$ stands for an interface between the two spinodally-decomposed phases in the bulk Fe–Mn system. Equation (10) gives a direct relation between $\kappa_X$, the interface energy and the free energy of the bulk material. The chemical free energy and excess enthalpy required for the current calculations are obtained from the ThermoCalc thermodynamic database TCFE9. Since the value for $\gamma_I$ is not known for our system, we examine here $\kappa_X$ values corresponding to $\gamma_I = 0.01$ to $0.1$ J m$^{-2}$. We then use the results from atom probe tomography (APT) measurements to adopt the best choice of $\gamma_I$ ($\kappa_X$) and GB density $\rho^{GB}$ in our studies.

## 3.2 Grain boundary segregation in the Fe–Mn system

### 3.2.1 Equilibrium Mn segregation

In order to address segregation in the Fe–Mn system, we study the equilibrium segregation isotherms, i.e. the GB equilibrium concentration $X_{Mn}^{GB}$ as a function of the bulk composition. In the following, we study Mn segregation in different Fe–Mn alloys annealed at 450 ºC. First parametric studies have been conducted to understand the effect of the GB atomic density $\rho^{GB}$ and the concentration gradient energy coefficient $\kappa_X$ values on the segregation isotherms. The results are shown in Fig. 2. The values with closest agreement to the APT measurements were then taken for conducting 3D simulations.

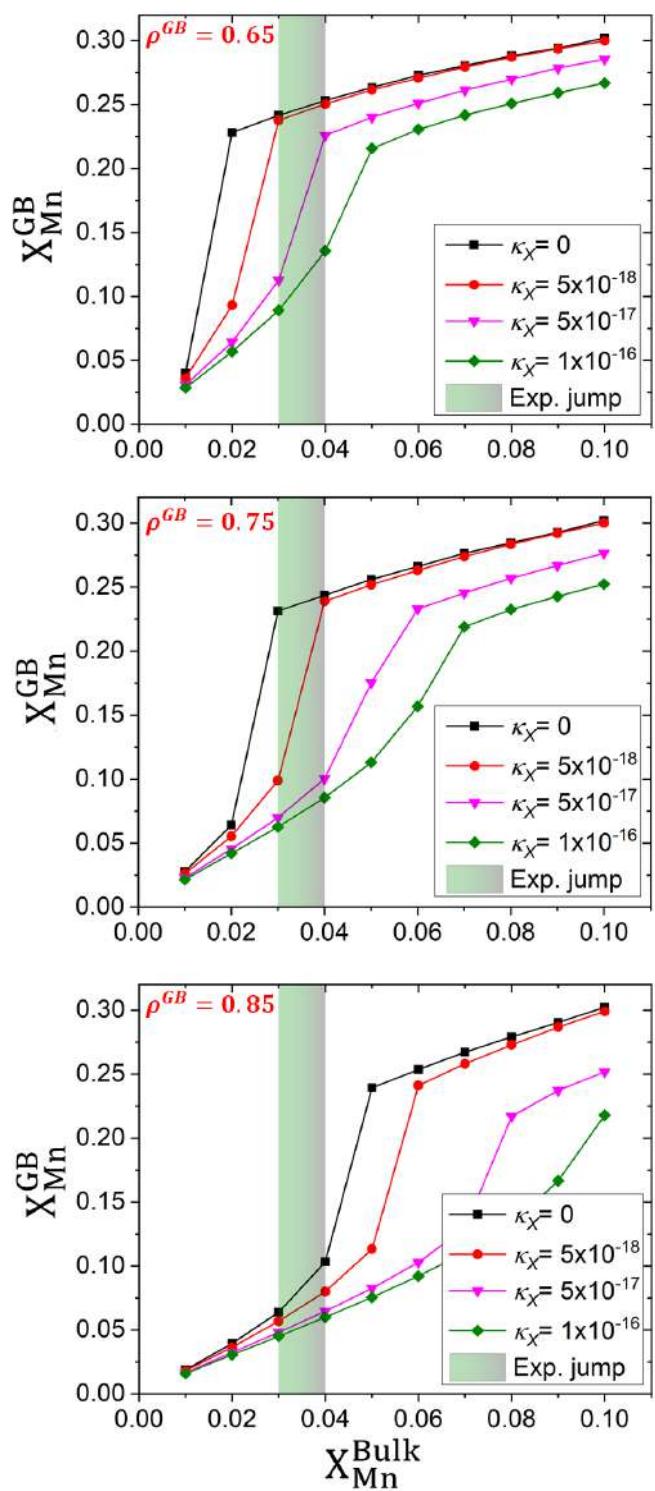

Figure 2: GB segregation isotherms. Three different GB atomic densities were examined, from top: $\rho^{GB}$ =0.65, 0.75 and 0.85. The colored area indicates where the GB spinodal was observed experimentally using APT analysis. The jump in the segregation isotherms denotes the interfacial spinodal. A good match between the APT results and simulations was obtained for a GB atomic density of 0.75 and a gradient coefficient $\kappa_x = 5 \times 10^{-18}$ J m² mol⁻¹.

The APT measurements revealed that at 450 °C a first order transition, marked by a distinct jump in the GB concentration, occurs for an alloy with a composition between 3.0 and 4.0 at.% Mn [38]. This range is marked in Fig. 2. The simulation results show that the segregation isotherm shifts to the left for a lower GB atomic density $\rho^{GB}$, i.e. the first order transition becomes possible for lower bulk compositions when the GB density decreases. Figure 2 also shows that higher levels of GB segregation can be achieved for a lower gradient coefficient $\kappa_X$. The optimal values (with the least deviation from the APT results) for $\rho^{GB}$ and $\kappa_X$ are found to be 0.75 and $5 \times 10^{-18}$ J m$^2$ mol$^{-1}$, respectively. These values are used for further investigations in the following.

Figure 3(a) shows the equilibrium GB concentration profiles obtained for different bulk compositions and by using $\rho^{GB} = 0.75$ and $\kappa_X = 5 \times 10^{-18}$ J m$^2$ mol$^{-1}$. The corresponding density profiles across the GB are shown in Fig. 3(d) and (e). While the GB width slightly increases in terms of the density variations, we found an abrupt increase in Mn segregation level, from about 8 to 22at.% Mn, when the bulk composition is changed by only 1at.%, namely from 3.0 to 4.0at.% Mn. The equilibrium Mn segregations are confined to the GB region. From the experiments, three BCC Fe–Mn alloys with 3.0, 4.0 and 8.6at.% Mn (in the following referred to as Fe3Mn, Fe4Mn and Fe9Mn respectively) were analyzed with respect to the Mn concentration (Fig. 3(b) and (c)).

In practice, segregation profiles across a GB can be asymmetric (Fig. 3(b)). This can be mainly due to the density difference of the adjoining atomic planes (with different atomic spacing) on the two sides of the GB. To account for this crystallographic feature, the current isotropic description of the atomic density field can be replaced with an anisotropic description. Broken bond models for GB segregation already suggest how to consider such asymmetry at the GB region [47, 48]. A similar idea can be adopted to introduce an anisotropic atomic density field. Nevertheless, the current simulation results based on the isotropic atomic density field show good agreement with the experiments as compred in Fig. 3(a)–(c).

The simulation results show that for a constant concentration gradient coefficient $\kappa_X$, a higher segregation level results in a wider (but still confined) segregation region. A further detailed study over the bulk composition space suggests that the first order transition in Mn segregation occurs for an alloy with a composition close to Fe–3.33at.% Mn in which the GB segregation level abruptly increases from $\sim$11 to $\sim$22.6at.% Mn (see Sec. 4).

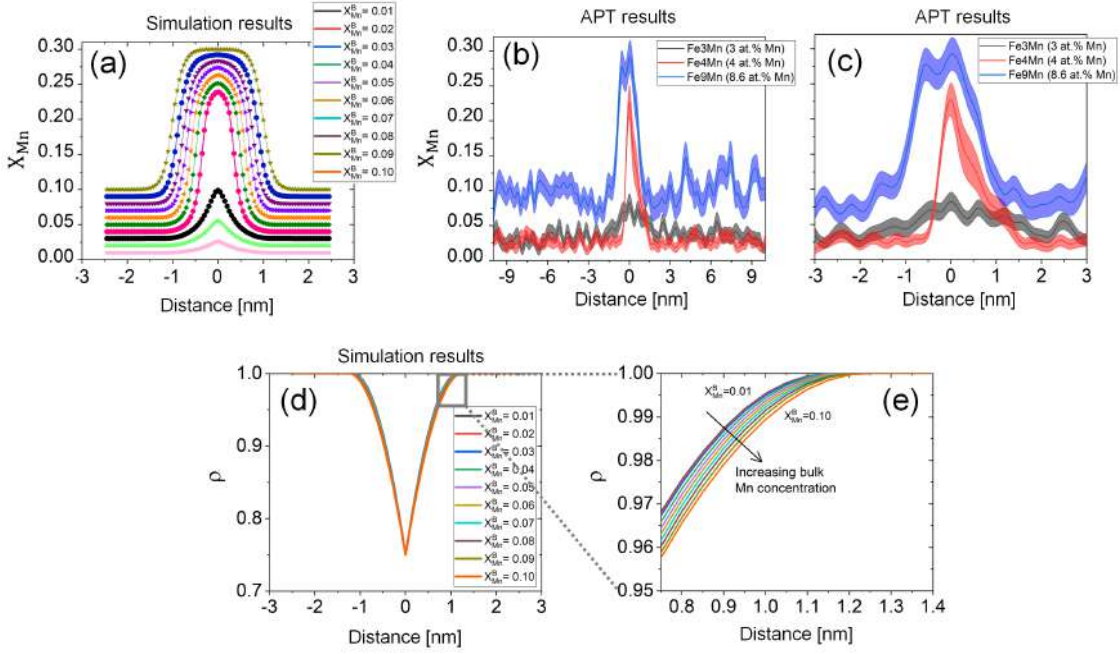

Figure 3: Equilibrium GB concentration and density profiles. (a) Mn concentration profiles are shown for different bulk composition (1D calculations). An atomic GB density $\rho^{GB} = 0.75$ and a gradient energy coefficient $\kappa_X = 5 \times 10^{-18}$ J m$^2$ mol$^{-1}$ were used in these calculations. (b) and (c) show the concentration profiles with corresponding error bars obtained from the APT measurements for three different Fe–Mn alloys with 3.0, 4.0 and 8.6at.% Mn (Fe3Mn, Fe4Mn and Fe9M, respectively). (b) is a magnification of (c) in x axis. A jump in the GB concentration occurs between 3.0 and 4.0at.% Mn that is also confirmed by APT measurements. The highest segregation level compares well in all three cases. (d) and (e) show equilibrium density profiles for different bulk compositions, corresponding to the concentration profiles in (a). For higher segregation levels, the GB width increases.

### 3.3 Segregation-induced spinodal and transient spinodal

While 1D calculations provide direct insight into equilibrium segregation isotherms, 3D simulations are required to study kinetics and patterns of the segregation to GBs. 3D simulations of GB segregation were conducted for three Fe–Mn alloys with 2.9, 3.3 and 9.0at.% Mn. The time evolution of the concentration field inside the GB plane is shown in Fig. 4. For an alloy with 2.9at.% Mn (below the critical composition), the simulations show that the GB segregation starts immediately and increases monotonically. The equilibrium GB concentration of ∼8at.% Mn is achieved which remains unchanged even after $2.4 \times 10^6$ s (∼ 28 days) at 450 $^o$C. At the same time, the initial concentration fluctuations, in the range of up to ±1at.%, decline and disappear. Similar results were obtained from the APT analysis for the Fe3Mn alloy annealed for 2 months at 450 $^o$C (Fig. 5 (a) and (b)): The GB is enriched with 8at.% Mn, very close to the simulation results. The APT analysis shows that concentration fluctuations up to ±3at.% exist inside the GB plane that remain stable even after the long-term annealing up to two months at 450 $^o$C. These are associated with the atomic density fluctuations that are naturally expected in a real GB, not reflected in the current atomic density-based model. Nevertheless, the concentration fluctuations are about an order of magnitude below the experimentally observed chemical spinodal fluctuations as discussed in the following.

For an alloy with 3.3at.% Mn (close to the critical composition), the simulations revealed an interfacial spinodal phase separation (Fig. 4). It was found that the initial GB segregation is followed by a gradual in-plane phase separation into low and high concentration domains with ∼8 and ∼22at.% Mn, respectively. Thereafter, islands with high-concentration level gradually grow and, above a critical size of ∼ 3 nm, start to coalesce and form larger segregation islands. The segregation kinetics of these later stages, however, are very slow. For the Fe4Mn alloy annealed for 2 month at 450 $^o$C, a jump in the segregation level was observed in the APT measurements (Figure 5 (c) and (d)) where the high and low concentration domains within the GB plane were observed next to each other.

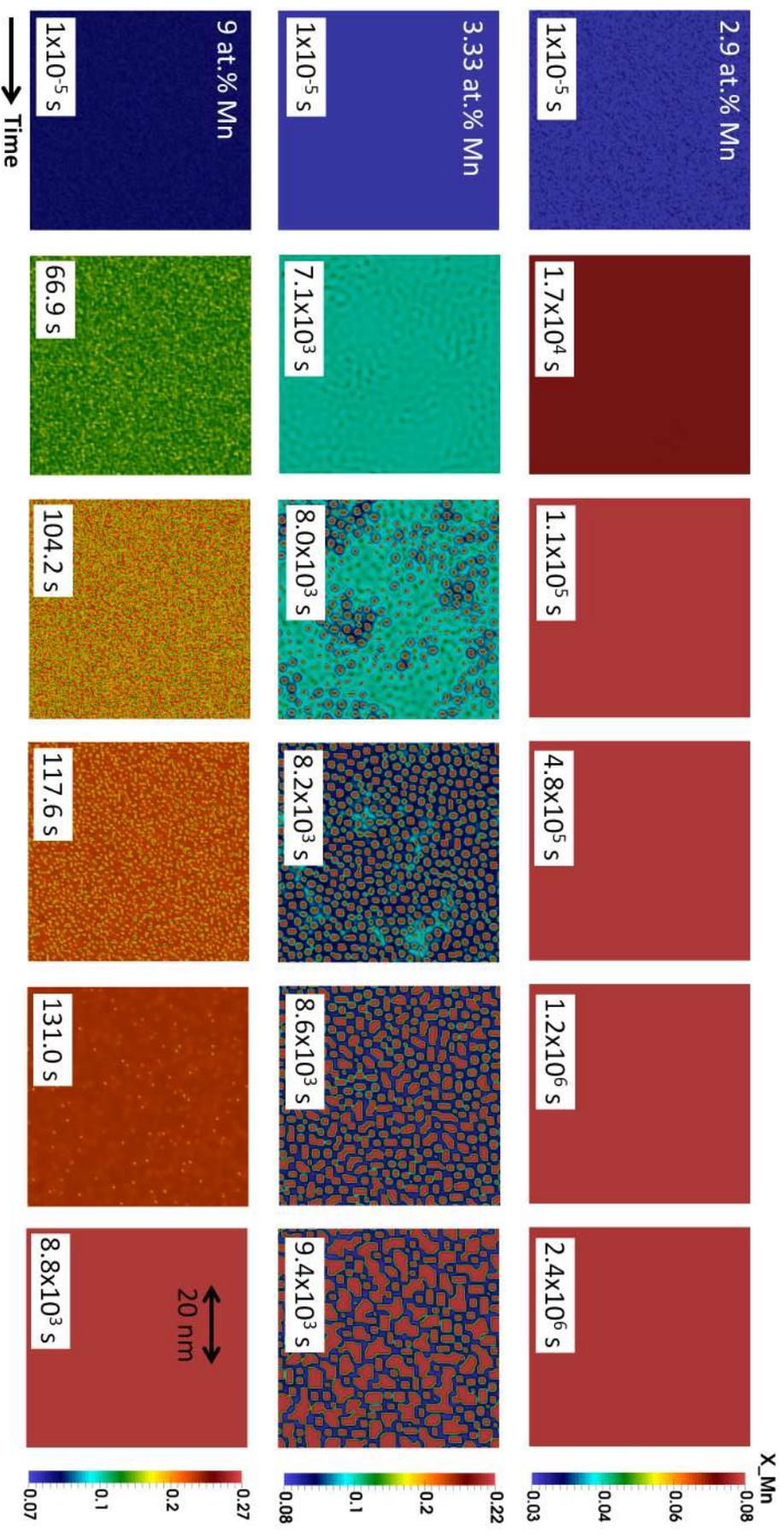

Figure 4: Time evolution of Mn concentration in the GB plane. 2D cross sections from 3D simulations are shown for three different alloys with 2.9, 3.3 and 9at.% Mn (450 °C). For an alloy with 3.3at.% Mn (close to critical composition) an interfacial spinodal phase separation is observed. Above this composition, we have found a transient spinodal regime demonstrated for Fe–9at.% Mn alloy.

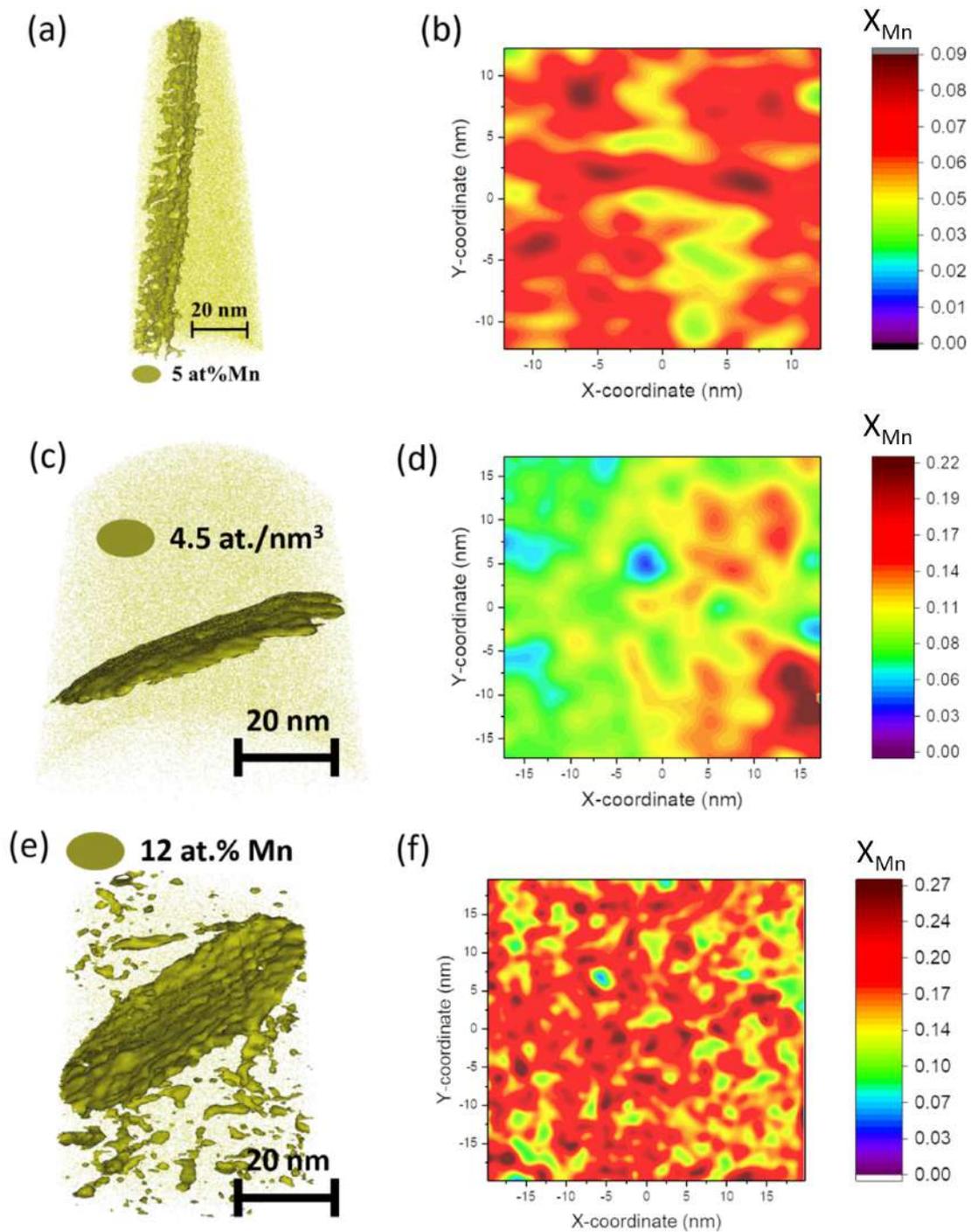

Figure 5: APT analysis of the Mn concentration. Three alloys with 3.0at.% Mn (a, b), 4.0at.% Mn (c, d), and 8.6at.% Mn (e, f) were studied. The 2D GB in-plane concentration maps are extracted from the 3D APT data. The Fe3Mn and Fe4Mn alloys were annealed for 2 months at 450 ºC. The Fe9Mn was annealed for 6h at the same temperature.

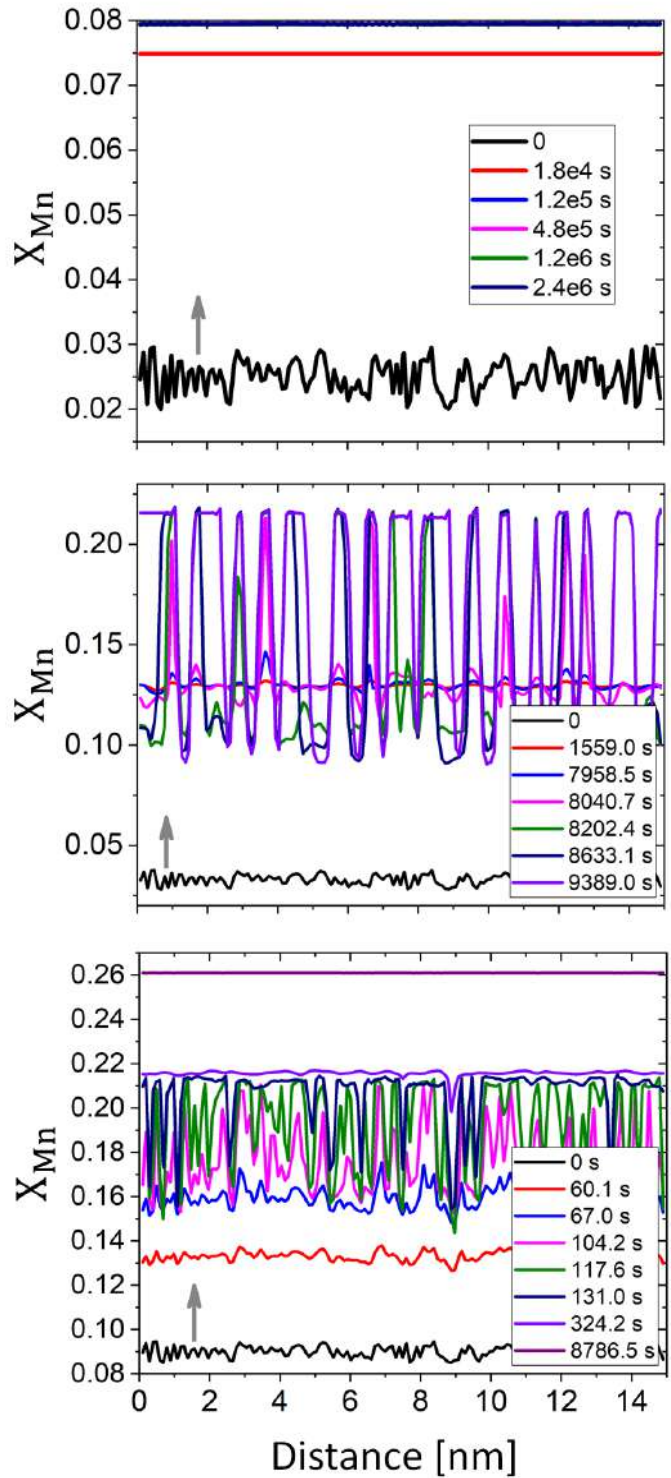

Figure 6: Simulated time evolution of the Mn concentration inside the GB plane: From top, 2.9, 3.3 and 9at.% Mn alloy compositions were studied. For the under-critical composition (2.9at.% Mn), segregation increases monotonically and it is stable even after long simulation times. For a composition close to the critical composition (3.3at.% Mn), a spinodal GB phase separation is observed. Above the critical composition (9at.% Mn) a transient spinodal regime is observed before reaching the equilibrium Mn segregation level.

The simulation results for a Fe–9at.% Mn alloy show an even more interesting segregation behavior. When approaching the GB spinodal transition region, it is found that the GB goes through a *transient* spinodal regime. In this regime the fluctuations in the composition grow and high-concentration regions with ∼22at.% Mn concentration form, which continues until the transitional spinodal phase separation is completed. In a final step, the segregation proceeds homogeneously until the equilibrium GB concentration is reached. The APT analysis confirms these numerical prediction: The Fe9Mn alloy annealed for 6 hours at 450 $^o$C shows a similar level of GB segregation with spatial fluctuation corresponding to a transient spinodal phase separation. The simulated time dependent composition along a line within the GB plane is presented in Fig. 6. The line-plots reveal the GB concentration and its fluctuations more clearly: For an alloy with 2.9at.% Mn, a smooth and flat GB concentration profile develops over time while a GB spinodal decomposition is clearly observed for Fe–3.3at.% Mn. A transient spinodal decomposition occurs in the Fe–9at.% Mn alloy. The current results show that the kinetics of GB segregation can be very complex not only close to the spinodal point but also for compositions above the spinodal composition. The concept of transient spinodal phase separation provides insights for material design purposes that will be discussed in the next section.

# 4   Discussion

Segregation of Mn has been considered as one possible cause for GB embrittlement in Fe–Mn alloys, which reduces the mechanical toughness of these alloys [76, 77]. This is attributed to GB decohesion due to the Mn segregation [20]. A recent DFT investigation indicates that a higher Mn concentration due to the segregation decreases the cleavage-fracture energy of the GBs [78]. If the Mn segregation level is high enough, it can initiate formation of austenite at the GBs that partly recovers the alloy toughness. Using transmission electron microscopy and near-atomic scale tomographic measurements, it was shown that reversed austenite layers can form on the Mn-enriched martensite boundaries in a Fe–9wt.% Mn alloy annealed at 450 $^o$C [27, 28]. The observed high levels of Mn segregation to the GBs is then attributed to a first-order segregation transition, i.e. a GB spinodal phase separation: Figure 7 shows the GB segregation isotherm for the Fe–Mn system at 450 $^o$C obtained from the current atomic density-based model. The jump in the segregation isotherm corresponds to an interfacial spinodal transition that confirms the abrupt increase in the GB segregation level observed in the experiments. This means that for the critical bulk composition (∼3.3at.% Mn) at 450 $^o$C a two-phase GB is expected to be in equilibrium with a single-phase bulk. Since, however, there is always a small deviation from the critical composition, a stable two-phase GB in equilibrium with the single-phase bulk can not be realized experimentally. The equilibrium concen-tration profiles across the GB are shown in Fig. 3 for different alloy compositions. The spread of the Mn segregation in the GB region is determined by the initial bulk composition as well as the concentration gradient coefficient $\kappa_X$. A larger gradient coefficient results in a wider and smoother segregation profile across the GB and reduces the maximum equilibrium segregation, i.e. it shifts the segregation isotherm towards lower values of GB segregation (Fig. 2). In systems with strong atomic interactions the concentration gradient coefficient can also be composition-dependent [79] rendering the spread of the segregation region composition-dependent as well.

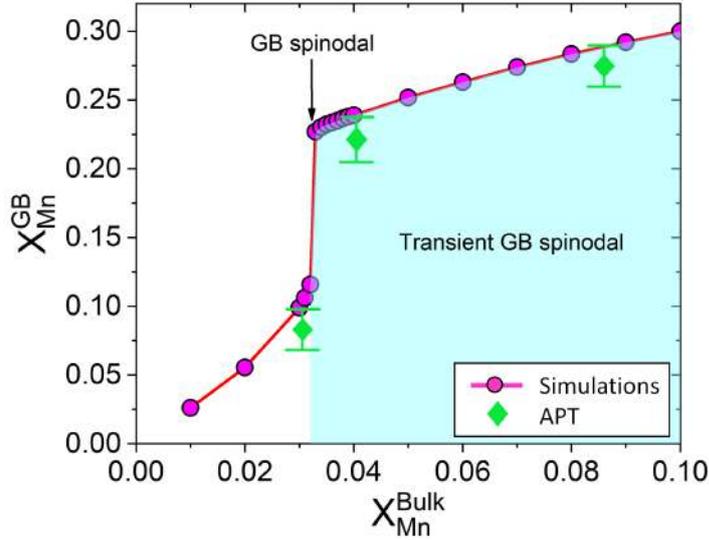

Figure 7: GB segregation isotherms (equilibrium GB concentration as a function of bulk composition) for the Fe–Mn system at 450 $^o$C. An atomic GB density $\rho^{GB} = 0.75$ and a gradient energy coefficient $\kappa_X = 5 \times 10^{-18}$ J m$^2$ mol$^{-1}$ were used for these calculations. The abrupt jump in the GB concentration (GB spinodal) is found for an alloy with ∼3.33at.% Mn. Above this composition a transient GB spinodal was revealed through 3D simulations. The green points indicate the highest GB concentration obtained from the APT measurements for alloys with 3.0, 4.0 and 8.6at.% Mn.

The interfacial spinodal point in the equilibrium segregation isotherm (Fig. 7) separates the low and high GB segregation regimes as a function of alloy composition. The results from our model show that the kinetics of segregation is very different for the low and high segregation levels. In particular, for bulk compositions above the interfacial spinodal composition, the kinetics of the GB segregation is found to be complex: The results of 3D simulations reveal that at 450 $^o$C and for alloy compositions $X^B > 0.033$, the GBs go through a transient spinodal regime before reaching a higher uniform Mn segregation level. Figures 4 and 6 show time evolution of the GB concentration for three Fe–Mn alloys. The transient GB spinodal is dictated by the fact that for reaching equal chemical potential (between the bulk and the GB) the segregation must proceed by passing through the GB spinodal area in the composition space. The chemical potential of the Mn solute atoms (relative to the Fe atoms) reads:

$$\delta\mu(T, \rho, X_{Mn}) = G^0_{Mn} - G^0_{Fe} + \rho^2 \frac{\partial \Delta H^B_{ex}}{\partial X_{Mn}} - T\frac{\partial \Delta S^B_{mix}}{\partial X_{Mn}} - \kappa_X \nabla^2 X_B. \qquad (11)$$

Using the thermodynamic data from ThermoCalc TCFE9 and neglecting the last term in this relation for simplicity, one can plot the chemical potential as a function of composition and for different GB atomic densities. Figure 8 (a) shows the chemical potential of Mn within the bulk ($\rho = 1$) and a GB with an atomic density $\rho = \rho^{GB} = 0.75$. It is found that above the GB spinodal and before reaching the bulk spinodal a range of bulk composition exists that produces a GB transient spinodal. The difference between the bulk and GB chemical potentials arises due to the enthalpy of mixing (third term in Eq. (11)) which quadratically scales with the local atomic density $\rho$. In the Fe–Mn system, the positive magnetic enthalpy of mixing plays an important role in the spinodal decomposition. The ranges of bulk composition and chemical potential for which a transient spinodal become possible are marked in Fig. 8.

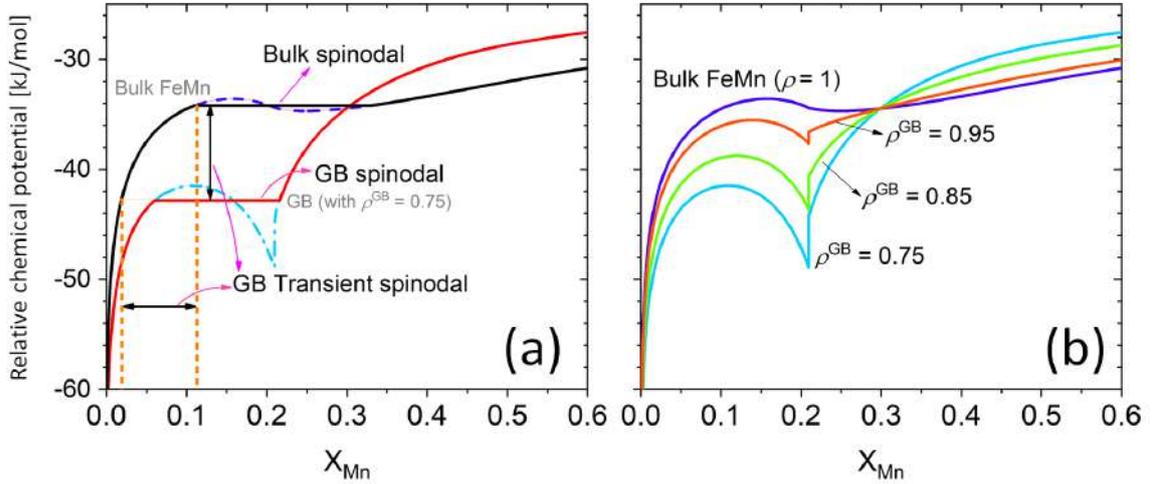

Figure 8: The chemical potential of Mn for bulk and different GBs. (a) The equilibrium chemical potentials of Mn for the bulk and a GB with average atomic density $\rho^{GB} = 0.75$ in Fe–Mn system at 450 ºC. Above the GB spinodal, segregation occurs with a transient spinodal phenomenon as discussed for the Fe–9at.% Mn alloy. (b) The chemical potentials for different types of GBs (represented here in terms of different GB atomic densities) are shown. Depending on the GB atomic density, the GB free energy density, the chemical potential and hence the coexistence of the bulk and GB change accordingly.

Obviously, different types of GBs with different structures and misorientations may show different atomic density profiles and average GB atomic densities $\rho^{GB}$ which determine the GB free energy and chemical potential in the current model approach. Figure 8 (b) shows the chemical potential of Mn for three different average GB atomic densities. At a given temperature, a higher GB atomic density value ($\rho^{GB} \to 1$) represents a GB that behaves more like the corresponding bulk. Hence, the composition/chemical potential window for a GB transient spinodal becomes smaller, as shown in Fig. 8 (b). Special GBs, e.g. coherent twin boundaries and highly symmetric coincidence site lattice boundaries, are expected to show higher average density values, close to the bulk density ($\rho^B = 1$) and therefore lower segregation. The characteristic GB density $\rho^{GB}$ can be associated with the average 'GB free volume'. Low-angle GBs are expected to show an average atomic density inversely proportional to their misorientation angle. In fact, it has been shown that the average free volume of the low-angle GBs increases as their misorientation angle increases [80]. A larger free volume is equivalent with a smaller GB atomic density. High-angle GBs (with the exception of spacial GBs) have the lowest average atomic density (highest free volume). In this case, a transient GB spinodal becomes more probable because the difference between the GB spinodal and the bulk spinodal chemical potentials increases, as depicted in Fig. 8.

Using available bulk thermodynamic data, the current atomic density-based model can provide quantitative understanding of GB segregation and phase separation for different types of GBs. The existence of a transient GB spinodal as revealed in the current study opens a novel route to design and tune desired precursor states for subsequent heterogeneous nucleation and phase transformation paths at GBs. In the Fe–Mn system, for instance, formation of reverse austenite at the Mn-enriched GBs plays a critical role in controlling GB embrittlement [27, 28, 81, 82]. Using the transient spinodal concept, several parameters of mechanical processing and heat treatment conditions can be adjusted to obtain desirable microstructures in alloys which are characterized by spatially confined spinodal and phase formation states. The fact that decoration to defects such as interfaces or dislocations enables these local thermodynamic phenomena allows to imprint site specific transformation effects into microstructures. While the initial alloy composition defines the thermodynamic feasibility and accessibility of a spinodal phase separation, the adequate thermo-mechanical processing can be applied to alter the GBs types and volume fraction in the system. In addition, the heat treatment conditions control the duration of the transient spinodal phase separation within the system, which strongly relates to thermally-activated solute diffusion properties of the bulk and GBs. A systematic study of these controlling parameters therefore will enable the exploration of the design space for segregation-assisted confined phase changes at lattice defects.

# 5  Summary

We have developed an atomic density-based model for studying the thermodynamics and kinetics of GB segregation and interfacial phase separation. A characteristic GB atomic density $\rho^{GB}$, corresponding to the type of GB, is obtained by atomistic tight-binding simulations. Using the current model, a thermodynamic description for GBs can be derived based on the available bulk thermodynamic data. Depending on the bulk composition, low and high levels of equilibrium Mn segregations were observed in the Fe–Mn system, separated by a segregation–assisted interfacial spinodal phase separation. Our studies reveal a transient spinodal phase separation regime for alloy compositions above the interfacial spinodal point. In the presence of the transient spinodal segregation, GB segregation and phase separation in three Fe–Mn alloys were consistently explained. The demonstrated quantitative understanding about the segregation and the transient spinodal phase separation at GBs provides a powerful means for achieving desirable microstructures by the knowledge-based variation of alloying and processing parameters.

# 6  Methods

**Full-field calculations:** In order to perform density-based full-field simulations, an OpenMP parallel C++ code was developed to solve Equations (6) and (7) numerically. A finite difference scheme with adaptive time stepping has been used. All calculations were done for $T = 450$ $^oC$ and assuming infinitely large bulk phases, i.e. a constant concentration boundary condition normal to the GB plane while other boundaries were periodic. The GB properties are obtained using atomistic simulations. We apply a coarse-graining scheme to obtain continuous density from the atomistic simulations, as discussed in the next section and Supplementer Information. We use $dx = 0.1$ nm, initial $dt = 10^{-5}$ s. In all simulations, uniform Mn concentrations were used with max. $\pm 1$at.% random fluctuations.

The thermodynamic data for the BCC Fe–Mn system (up to 30 at.% Mn) were obtained from ThermoCalc TCFE9 and MOB04 databases. Physical parameters are presented in Table 1. All quantities are scaled by the BCC Fe molar volume to obtain respective values per unit volume. For the Mn atoms, the composition-dependent mobility from ThermoCalc database was fitted as $M_{Mn} = (1.3993 \times 10^{-26})e^{19.0375\, X_{Mn}}$ m$^2$ mol J$^{-1}$ s$^{-1}$ that indicates an increase in the atomic mobility as the Mn content increases (see also [67–72]). The simulation results are extracted and visualized using Paraview [83].

| Parameter | Values | Physical dimension | Source |
|---|---|---|---|
| $E_A^B$ (for BCC Fe) | $-4479.4$ | J mol$^{-1}$ | Atomistic Simulations |
| $S_A^B$ (for BCC Fe) | $53.1$ | J K$^{-1}$ mol$^{-1}$ | ThermoCalc database |
| $\rho^{GB}$ | $0.65 - 0.85$ | -- | Atomistic Simulations |
| $\eta_A$ | $0.6 - 1.5 \times 10^{-9}$ | m | Atomistic Simulations |
| $\kappa_\rho$ | $1.63 - 10.2 \times 10^{-16}$ | J m$^2$ mol$^{-1}$ | Model (Eq. (9)) |
| $\gamma_A$ | $1.28$ | J m$^{-2}$ | Atomistic Simulations |
| $\kappa_X$ | $0.05 - 1 \times 10^{-16}$ | J m$^2$ mol$^{-1}$ | Model (Eq. (10)) |
| $M$ | $1.34 \times 10^{-26} e^{19.0 X_{Mn}}$ | mol m$^2$ J$^{-1}$ s$^{-1}$ | ThermoCalc |
| $L$ | $1 \times 10^{-7}$ | mol J$^{-1}$ s$^{-1}$ | -- |
| $V_m^{BCC\,Fe}$ | $7.2 \times 10^{-6}$ | m$^3$ mol$^{-1}$ | ThermoCalc |

Table 1: Input parameters for the current simulation studies.

**Atomistic calculations:** In order to obtain realistic values for the GB atomic density $\rho^{GB}$, explicit atomistic simulations have been performed within the environmental tight-binding approach [73, 74]. This approach enables a fully quantum-mechanical parameter-free description of the energetics and forces of systems of arbitrary chemical complexity, while remaining sufficiently efficient so as to examine a broad variety of microstructural defects. In the present case, we consider a *tilt* GB, namely the $\Sigma 9\{122\}[1\bar{1}0]$ symmetric tilt GB in $\alpha$-Fe. A 144-atom supercell for this GB has been generated, and the structural parameters and internal coordinates have been fully relaxed within the tight-binding method. The resulting atomistic structure is illustrated in Fig. 9.

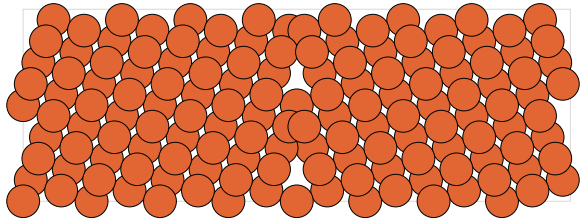

Figure 9: The structure of the $\Sigma 9$ $\{122\}[1\bar{1}0]$ symmetric tilt GB in $\alpha$-Fe.

From the atomistic simulation, we calculate the GB energy density $\gamma_A$. However, in order to obtain the atomic density field $\rho^{GB}$, we have to establish a connection between the discrete atomic structure of the GB and the continuous atomic density function as introduced in the current atomic density-based phase-field model. In the present case, this is done by a coarse-graining step. We replace the atomistically-obtained density function

$$\rho(r) = \sum_I \delta(r - R_I) \qquad (12)$$

with $R_I$ being the set of positions of the atoms, with a smeared-out density function, where the delta functions of Eq. (12) are replaced by a normalized Gaussian

$$\delta(x) = \frac{e^{-\frac{x^2}{2\beta^2}}}{\sqrt{2\pi}\,\beta} \qquad (13)$$

with a prescribed coarse-graining length $\beta$, so that the continuous atomic density profile becomes a smooth function in real space. The parameter $\beta$ should be at least of the order of the interatomic spacing ($a \sim 2.5$Å) in the material. A physically-motivated coarse-graining radius is the cut-off radius ($\sim$4 Å) over which the atomistic forces are calculated. A higher value of $\beta$ results in a smoother atomic density profile, but with the possible cost of being unable to resolve certain features of the GB itself. We examine three choices of the parameter $\beta$ close to the cut-off radius as discussed in Sec. 3.1.

**Experiments and APT analysis:** Three binary Fe–Mn alloys, identified with 3.0, 4.0, 8.6at.% Mn, referred to as Fe3Mn, Fe4Mn and Fe9Mn, were cast into a rectangular billet in a vacuum induction furnace. The composition of the alloys is shown in Table 2 according to wet chemical analysis. The slabs were hot-rolled at 1100 $^o$C from 60 to 6 mm thickness and then water quenched. Subsequently, highly segregated edges of the slab were cut off. The billets were reheated to 1100 $^o$C for 1 hour and water quenched to room temperature to minimize Mn banding. After water quenching from the homogenizing temperature the alloys were fully ferritic without retained austenite. The mechanisms of transformation from austenite to ferrite were martensitic transformation for the Fe9Mn alloy and massive transformation for the Fe3Mn and Fe4Mn alloys. The Fe9Mn was annealed for 6 hours, while the two other alloys were subsequently annealed up to 2 months at 450 $^o$C in order to characterize the equilibrium amount of segregation at the GBs. The Fe9Mn and Fe4Mn alloys are situated in the two-phase region of the phase diagram (ferrite and austenite are stable phases). The Fe3Mn alloy is situated in the single phase field of the phase diagram (ferrite is the only stable phase).

| Alloy | Mn | C | Ni | Al | S | P | O | N | Fe |
|---|---|---|---|---|---|---|---|---|---|
| Fe9Mn | 8.46 (8.60 at.%) | 0.0075 | 0.0175 | <0.002 | 0.0047 | <0.002 | 0.0102 | 0.0040 | balance |
| Fe4Mn | 3.92 (4.0 at.%) | 0.0081 | 0.0172 | <0.002 | 0.0036 | <0.002 | 0.0096 | 0.0071 | balance |
| Fe3Mn | 2.95 (3.0 at.%) | 0.0079 | 0.0214 | <0.001 | 0.0028 | <0.002 | 0.0196 | 0.0024 | balance |

Table 2: The composition of the alloys studied in this work.

APT specimens with end radii below 100 nm were prepared using a FEI Helios NanoLab600i dual-beam Focused Ion Beam (FIB)/Scanning Electron Microscopy (SEM) instrument. APT was performed using a LEAP 5000 XS device by Cameca Scientific Instruments, with approx. 80% detection efficiency, at a set-point temperature of 50 K in laser-pulsing mode at a wavelength of 355 nm, 500 kHz pulse repetition rate and 30 pJ pulse energy. For reconstructing 3D atom maps, visualization and quantification of segregation the commercial software IVAS by Cameca was employed following the protocol introduced by Geiser *et al.* [84] and detailed in Gault *et al.* [85]. The 3D-mapping was obtained by the Voltage-based reconstruction of the detected ions. The reconstructions were calibrated by the interplanar distance of the crystallographic planes associated with the low-hit density poles.

# Acknowledgements

Reza Darvishi Kamachali gratefully acknowledges financial support from the German Research Foundation (DFG) under the project *DA 1655/2-1* within the Heisenberg programme. A. Kwiatkowski da Silva is grateful to the Brazilian National Research Council (Conselho Nacional de Pesquisas, CNPQ) for the scholarship through the Science without Borders Project (203077/2014-8).

# Data availability

The numerical and experimental data from this study are available upon reasonable request.

# Author contributions

R.D.K. developed the density-based model, performed the simulations and wrote the manuscript; A.K.S. performed the experiments and APT studies; A.K.S., D.R., D.P. and B.G. conceived and supervised the experiments and APT analysis; E.M. conducted the atomistic simulations; J.N. supervised and discussed the simulation results; All authors reviewed the manuscript.

# Competing interests

The authors declare no competing interests.

Supplementary Information for

# Segregation—Assisted Spinodal and Transient Spinodal Phase Separation at Grain Boundaries

Reza Darvishi Kamachali\*, Alisson Kwiatkowski da Silva, Eunan McEniry, Dirk Ponge, Baptiste Gault, Jörg Neugebauer, Dierk Raabe

**Derivation of the Density-Based Gibbs Free Energy Functional**

In order to obtain the grain boundary Gibbs free energy as a function of its atomic density variation, we need to first formulate the general form of Gibbs free energy functional in terms of the atomic density field. In a pure substance made of atom $A$ at constant $T$ and $p$, the Gibbs free energy functional can be written as

$$\mathcal{G}_A = \int_\Omega G_A \, \mathrm{d}V = \int_\Omega \rho_n \, (H_A - TS_A) \, \mathrm{d}V \qquad \text{S.1}$$

where $\rho_n(\vec{r})$ is the atomic density field equivalent to the inverse molar volume field $V_m^{-1}(\vec{r})$. $H_A$ is enthalpy per unit mole with $H_A = K_A + E_A + pV_A$ ($K_A$: kinetic energy and $E_A$: potential energy, $V_A$: volume), and $S_A$ is entropy per unit mole, respectively. As van der

---

\*Corresponding Author, Email: kamachali@mpie.de

Waals suggested, the potential energy can be a function of the nonlocal interactions [1]. Assuming a planar grain boundary separating two infinitely large homogeneous grains, we can write the potential energy density as

$$E_A(x) = E_A(-\infty) + \frac{1}{2}\int_{-\infty}^{x} f(r)\,dr. \qquad \text{S.2}$$

In our system, breaking the symmetry only normal to the grain boundary plane, the forces can be realized considering only one spatial dimension. Here $E_A(-\infty)$ is the potential energy inside homogeneous grain at $r = -\infty$ and the integral describes the work done on a given amount of matter to bring it from $r = -\infty$ to a given position $r = x$. $f(r)$ is the sum of all forces acting on any point $r$:

$$f(r) = \int_{0}^{\infty}[\zeta(r+q) - \zeta(r-q)]\,dq \qquad \text{S.3}$$

where $\zeta(r \pm q)$ is the force density at point $r$ due to interaction between the two material points separated by a distance $\pm q$. The interaction forces between atoms depend on their interatomic potential $U(r \pm q)$ with $\zeta(r \pm q) = \frac{\partial U(r \pm q)}{\partial q}$. Although the detailed form of the interatomic potentials/forces can vary for different types of atoms, it is well-known that the atomistic interactions (in the absence of long-range electrostatic interactions which would be treated separately) rapidly decreases over the distance between atoms. For the sake of our discussion, we consider here a simple functional form for the interaction force as $\zeta(r \pm q) = \frac{\alpha\, n(r \pm q)}{q^z} = \psi(q)\,\rho_n(r \pm q)$ where $\alpha$ is a material constant, $n$ is the number of atoms, $\psi(q) = \frac{\alpha V_L}{q^z}$ and $z$ is a positive exponent. Such a relation is inspired by interatomic force relations with $z \gg 1$ [4]. In this formulation, $V_L$ is the characteristic coarse-graining volume of the current mesoscale description over which the atomic density field is measured as $\rho_n(r) = \frac{n(r)}{V_L}$. Similar coarse-graining descriptions are worked out in defining different mesoscale free energy formulations [2, 3].

In the theories of atomistic simulations the atomic forces are practically calculated up to a cut-off radius $R_c$ above which the atomic forces are neglected [4]. Hence one reasonable approximation for the coarse-graining volume is $V_L \approx \frac{4\pi}{3}R_c^3$ which relates the atomistic simulation length-scale with the current mesoscale density-based formulation. Furthermore, the specific form for $\psi(q)$ can be obtained from atomistic simulations as well. Inserting the force density $\zeta(r \pm q) = \psi(q)\rho_n(r \pm q)$ in Eq. S.3, and integrating by part we can write

$$f(r) = \int_0^\infty \psi(q)[\rho_n(r+q) - \rho_n(r-q)]\,dq$$

$$= \psi(q)[\rho_n(r+q) - \rho_n(r-q)]\Big|_0^\infty - \int_0^\infty \psi(q)\frac{\partial[\rho_n(r+q) - \rho_n(r-q)]}{\partial q}\,dq \qquad \text{S.4}$$

$$= -\int_0^\infty \psi(q)\frac{\partial[\rho_n(r+q) - \rho_n(r-q)]}{\partial q}\,dq$$

where $\lim_{q\to\infty}\psi(q) = 0$ and $\lim_{q\to 0}[\rho_n(r+q) - \rho_n(r-q)] = 0$ were considered. Applying Tylor expansions $\rho_n(r\pm q) = \rho_n(r) \pm \frac{\partial \rho_n}{\partial r}q + \frac{1}{2!}\frac{\partial^2 \rho_n}{\partial r^2}q^2 \pm \frac{1}{3!}\frac{\partial^3 \rho_n}{\partial r^3}q^3 + \cdots$ in Eq. S.4 gives

$$f(r) = -\int_0^\infty \psi(q)\left[2\frac{\partial \rho_n}{\partial r} + \frac{\partial^3 \rho_n}{\partial r^3}q^2 + \cdots\right]dq$$

$$f(r) = -A\left[\frac{\partial \rho_n}{\partial r}\right] - \kappa\left[\frac{\partial^3 \rho_n}{\partial r^3}\right] - \cdots$$

S.5

with two materials parameters $A_A = \int_0^\infty 2\,\psi(q)\,dq$ and $\kappa_A = \int_0^\infty \psi(q)q^2\,dq$. Using Eqs. S.2 and S.5 with the boundary values $E_A(-\infty) = -\frac{A_A}{2}\rho_n(-\infty)$ and $\left(\frac{\partial^2 \rho_n}{\partial r^2}\right)_{-\infty} = 0$ we obtain the potential energy as a function of density

$$E_A(x) = -\frac{A_A}{2}\rho_n(x) - \frac{\kappa_A}{2}\left(\frac{\partial^2 \rho_n}{\partial r^2}\right)_x - \cdots \qquad \text{S.6}$$

As a consequence of the Tylor expansion, it is clear that the fourth and higher order spatial derivatives of the atomic density field can still contribute to the potential energy. For the sake of simplicity, however, these higher order contributions are omitted in the following. Using Eq. S.6, the free energy functional (Eq. S.1) can be written as

$$\mathcal{G}_A = \int_\Omega \left[-\frac{A_A}{2}\rho_n^2 + \rho_n(K_A + pV_A - TS_A) + \frac{\kappa_A}{2}(\nabla \rho_n)^2\right]dV \qquad \text{S.7}$$

where we used a three dimensional notation and $\int (\nabla \rho_n)^2 \, dV = -\int \rho_n \nabla^2 \rho_n \, dV$ with $\nabla \rho_n = 0$ at the boundaries of the integral. As a result of the current derivation, two density terms with quadratic and linear scalings appear in Eq. S.7. This is further discussed in the main manuscript of our study. Equation S.7 gives the Gibbs free energy functional of a monatomic single phase system with spatial atomic density variations.

The atomic density variations on the two sides of the grain boundary can be different, depending on the crystallographic planes that meet at the grain boundary plane. This can result in asymmetric grain boundary structure and segregation. In the current study, however, we neglect this effect and assume the same bulk density on the two sides of the grain boundary. Thus we can further simplify our description by normalizing the density field with the corresponding bulk density value $\rho_n(-\infty) = \rho_n^B = (V_m^B)^{-1}$. This gives a relative density field $\rho(x) = \frac{\rho_n(x)}{\rho_n^B}$ and hence the general Gibbs free energy density can be written as

$$G_A = -\frac{A_A^B}{2}\rho^2 + \rho \left( K_A^B + pV_A^B - TS_A^B \right) + \frac{\kappa_\rho}{2} (\nabla \rho)^2 \qquad \text{S.8}$$

in which $A_A^B = A_A(\rho_n^B)^2$, $\kappa_\rho = \kappa_A(\rho_n^B)^2$, $K_A^B = K_A \rho_n^B$, $V_A^B = V_A \rho_n^B$, $S_A^B = S_A \rho_n^B$ and the Gibbs free energy of the homogeneous bulk phase (for $\rho = 1$) is

$$G_A^B = -\frac{A_A^B}{2} + K_A^B + pV_A^B - TS_A^B. \qquad \text{S.9}$$

Thus, subtracting S.9 from S.8 we obtain

$$G_A^{GB} = E_A^B(\rho^2 - 1) + C_A^B(\rho - 1) + \frac{\kappa_\rho}{2} (\nabla \rho)^2 \qquad \text{S.10}$$

where we defined $E_A^B = -\frac{A_A^B}{2}$ and $C_A^B = K_A^B + pV_A^B - TS_A^B$. Equation S.10 gives an approximation of the grain boundary Gibbs free energy density $G_A^{GB}$. This is the starting point in our model development which is further discussed in the main manuscript of the current study.